\documentclass[useAMS,usenatbib,letterpaper,onecolumn]{mn2e} 
\usepackage[totalwidth=480pt,totalheight=680pt]{geometry}
\usepackage{graphicx,amssymb,amsbsy,amsmath}
\newcommand{\be}{\begin{equation}}
\newcommand{\ee}{\end{equation}}
\def\lta{\,\raise 0.3 ex\hbox{$ < $}\kern -0.75 em
 \lower 0.7 ex\hbox{$\sim$}\,}
\def\gta{\,\raise 0.3 ex\hbox{$ > $}\kern -0.75 em
 \lower 0.7 ex\hbox{$\sim$}\,}

\newcommand{\mtot}{m_{\scriptstyle T}} 
\newcommand{\mplan}{ m_{\rm p}} 
\newcommand{\mcrit}{m_{\scriptstyle C}} 
\newcommand{\nump}{ {\cal N}_{\rm p} } 
  
\newcommand{\selfg}{\alpha_{\rm g}}

\newcommand{\rplan}{R_{\rm p}} 

\title[Energy Optimization in Extrasolar Planetary Systems] 
{Energy Optimization in Extrasolar Planetary Systems: \\
The Transition from Peas-in-a-Pod to Runaway Growth} 

\author[Adams, Batygin, Bloch, Laughlin]{Fred C. Adams,$^{1,2}$ 
Konstantin Batygin,$^3$ Anthony M. Bloch,$^4$ Gregory Laughlin$^5$\\  
\\ 
$^1$Physics Department, University of Michigan, Ann Arbor, MI 48109\\
$^2$Astronomy Department, University of Michigan, Ann Arbor, MI 48109\\
$^3$Division of Geological and Planetary Sciences, 
California Institute of Technology, Pasadena, CA 91125\\
$^4$Math Department, University of Michigan, Ann Arbor, MI 48109\\
$^5$Astronomy Department, Yale University, New Haven, CT 06511}

\begin{document} 

\date{$\quad$ November 2019} 

%\pagerange{\pageref{firstpage}--\pageref{lastpage}} \pubyear{2020}
%\onecolumn                                                                     
\maketitle

\label{firstpage}

\begin{abstract} 
Motivated by the trends found in the observed sample of extrasolar
planets, this paper determines tidal equilibrium states for forming
planetary systems --- subject to conservation of angular momentum,
constant total mass, and fixed orbital spacing. In the low-mass limit,
valid for superearth-class planets with masses of order
$\mplan\sim10M_\oplus$, previous work showed that energy optimization
leads to nearly equal mass planets, with circular orbits confined to a
plane. The present treatment generalizes previous results by including
the self-gravity of the planetary bodies. For systems with
sufficiently large total mass $\mtot$ in planets, the optimized energy
state switches over from the case of nearly equal mass planets to a
configuration where one planet contains most of the material. This
transition occurs for a critical mass threshold of approximately
$\mtot\gta\mcrit\sim40M_\oplus$ (where the value depends on the
semimajor axes of the planetary orbits, the stellar mass, and other
system properties). These considerations of energy optimization apply
over a wide range of mass scales, from binary stars to planetary
systems to the collection of moons orbiting the giant planets in our
solar system.
\end{abstract} 


\section{Introduction} 
\label{sec:intro} 

Thousands of extrasolar planets have been discovered over the past two
decades. Although the observed collection of planetary systems
exhibits a wide range of properties \citep{borucki,batalha}, a
significant subset of the extrasolar multi-planet systems display an
apparent but unexpected degree of regularity: The planets within these
systems tend to have nearly equal radii \citep{weissa,weissb} and
nearly equal masses \citep{millholland,songhu}. Taken together, these
findings jointly suggest similar mean densities (and perhaps similar
chemical compositions). Beyond the physical characteristics of the 
planets themselves, the planetary orbits are observed to be nearly
co-planar \citep{tredong,fangmargot} with low eccentricities
\citep{vaneylen,mills2019} and uniform spacing \citep{rowe,steffen}. 
This striking intra-system uniformity (see also \citealt{weisspet})
holds for planet masses of order $\mplan\sim3-10M_\oplus$, the typical
mass for these bodies \citep{zhu}, but breaks down for systems
containing giant planets \citep{songhu}. A fundamental goal of planet
formation theory is to understand how planetary systems attain their
final properties.  In particular, we would like to identify the
preferred configurations of these systems and the basic physical
principles that shape these outcomes.

Previous work indicates that the aforementioned order can be
understood, in part, through a process of energy optimization that
operates while the planets are being assembled \citep{adams2019}.
Specifically, forming pairs of planets are assumed to have constant
total angular momentum ${\bf L}$, constant total mass $\mtot$, and a
given orbital spacing $\Lambda=a_2/a_1$ (where $a_j$ are the semimajor
axes). With these constraints, the lowest energy state accessible to a
forming two-planet system has nearly equal mass planets, along with
vanishing orbital eccentricities $e_j=0$ and mutual inclination $i=0$.
This extremization procedure can be extended to systems with larger
numbers of planets. Under the assumption that energy optimization
operates on adjacent planetary pairs (see \citealt{adams2019} for
further detail), the resulting solar system properties are roughly
consistent with current observations.  Moreover, this result depends
only on the lowest energy state available to the system and is
independent of any particular paradigm for planet formation.

The goal of this paper is to generalize previous calculations of
energy optimization during the planet assembly process by including
the self-gravitational potential of the planetary bodies. The
well-ordered planetary systems described above generally have
super-earth masses \citep{zhu}. For systems that contain more massive
Jovian planets, however, the masses, orbital spacing, and other
properties are not nearly as uniform \citep{songhu}. Moreover,
previous work shows that the self-gravity of the forming planets can
be a significant component of the energy budget for sufficiently large
bodies (e.g., see Figure 3 of \citealt{adams2019}). Motivated by these
prior results, this paper includes self-gravity in the energy
optimization analysis. For low-mass systems, we find that planetary
systems are optimized with mass fractions $f\sim1/2$, corresponding to
nearly equal mass planets. However, when the total mass of forming
planetary pairs becomes sufficiently large, $\mtot>\mcrit$, the
optimal mass fraction $f\to0$, which implies that it becomes
energetically favorable for one of the planets to experience runaway
growth (so that $m_1\to0$ and $m_2\to\mtot$). The mass threshold is
found to be $\mcrit\sim40M_\oplus$ for $a\sim0.1$ AU, but displays a
significant dependence on the semimajor axis $a_2$ of the remaining
planet, as well as the orbital spacing parameter $\Lambda$ and stellar
mass $M_\ast$.

Energy optimization calculations, subject to constraints such as
conservation of angular momentum, have a long history in astrophysics
(beginning with \citealt{darwin1,darwin2}). Previous examples of such
Darwin problems include the tidal equilibrium states for binary star
systems \citep{counselman,hut1980}, stable configurations for Hot
Jupiter systems \citep{levrard,ab2015}, and hierarchical
star-planet-moon systems \citep{ab2016}. In these previous
applications, the masses of the constituent bodies were specified and
held constant. In previous work \citep{adams2019}, we considered a new
type of Darwin problem for which the masses of the individual planets
are variable, but the total mass is constant. The mass of the planets
is thus apportioned in order to achieve the lowest energy state. This
paper generalizes the problem further by including the self-gravity of
the planetary bodies in the energy budget.

In this class of problems, one assumes that the physical system
attains its lowest energy state through some type of dissipation,
whose existence is implicitly assumed but is not included explicitly.
By stepping back from the modeling of detailed --- but often
unobservable --- microphysics, one can make considerable progress
toward identifying the preferred end states. In the case of binary
stars, for example, the optimized final state has the stellar rotation
rates synchronized with the orbital frequency, all three angular
momentum vectors pointing in the same direction, and zero eccentricity
\citep{hut1980}.  Moreover, the existence and properties of this
optimized end state are independent of the evolutionary trajectories
by which it can be attained.\footnote{Nonetheless, the optimal state 
is not always realized. In the case of binary stars, for example, 
the tidal dissipation process can take longer than the ages of the 
systems. } Similarly, the original example \citep{darwin1} explained 
the dynamics of Earth-Moon tidal interactions at a sophisticated
quantitative level while giving only a qualitative description of 
the dissipation mechanism. The goal of the present paper is thus to 
specify the properties of the optimal configurations for multi-planet
systems in analogous fashion.

\section{Energy Optimization for Planetary Pairs} 
\label{sec:model}  

This section performs the energy optimization procedure for a system
of two planets orbiting their host star. The angular momentum and
orbital spacing are held fixed. The total mass in planets is also
constant, but the mass in each planet is allowed to vary. The 
general formulation of the problem is presented in Section
\ref{sec:formulate}, which includes the self-gravity of the planets.
The optimized state is found in Section \ref{sec:dfirst}, where we
also identify a mass threshold that separates systems with nearly
equal mass planets from those in which one planet dominates the mass
budget. Section \ref{sec:dsecond} considers the second variation and
shows that the extremal solution represents a minimum of the total
system energy.
 
\subsection{Formulation of the Problem} 
\label{sec:formulate} 

In this incarnation of the Darwin problem, the energy budget includes
both the orbital energy of the planets and their self-gravity. The
total system energy can thus be written in the form 
\be
{\cal E} = - {G M_\ast m_1 \over 2a_1} - {G M_\ast m_2 \over 2a_2} 
- \selfg {G m_1^2 \over R_1} - \selfg {G m_2^2 \over R_2} \,,
\ee
where $\selfg$ is a dimensionless factor of order unity and is
determined by the internal structure of the planets. For simplicity,
the constant $\selfg$ is assumed to be the same for both planets. 
The planets have masses $m_j$, radii $R_j$, and orbits with semimajor
axes $a_j$; the host star has mass $M_\ast$.  Note that this treatment
ignores the planet-planet interaction energy, which is much smaller
than the above terms (see \citealt{adams2019} for further discussion).

In general, the orbits of the planets are characterized by elements
$(a_j,e_j,i)$, where $i$ is the angle between the normal directions of
the orbits. However, one can show that the optimization procedure
requires the eccentricities to vanish ($e_j=0$) and the mutual
inclination $i\to0$. This result can be understood as follows: The
angular momentum for this problem is the same as that used previously,
and the new energy terms only depend on the planetary masses $m_j$.
As a result, the optimization procedure results in zero eccentricities
and co-planar orbits, and all of the mixed derivatives in the second
variation are zero (see equations [31--36] of \citealt{adams2019}).
We can thus set $e_j=0$ and $i=0$, so that only the magnitude 
${\cal L}$ of the angular momentum must be considered. Notice also
that resonant as well as secular planet-planet interactions are
reduced when $e_j\to0$ and $i\to0$, so our neglect of the disturbing
function is consistent \citep{md}.

The total angular momentum ${\cal L}$ can be written in the form
\be
{\cal L} = m_1 (G M_\ast a_1)^{1/2} + m_2 (G M_\ast a_2)^{1/2} \,, 
\ee
where this expression is valid in the limit $m_j \ll M_\ast$ (keep 
in mind that $m_j\sim10M_\oplus\sim3\times10^{-5} M_\ast$). Although 
the individual masses are allowed to vary, the total mass contained in
planetary bodies is held constant so that 
\be
m_1 + m_2 \equiv \mtot = constant\,. 
\ee 
The orbital spacing is also considered fixed, with a spacing parameter 
defined by 
\be
\Lambda \equiv {a_2 \over a_1} = constant \,.
\ee
By convention we take $a_2>a_1$ so that $\Lambda>1$. 

The motivation for keeping $\Lambda$ constant arises both from
observations and dynamical considerations. First we note that gravity
is scale-free, which is consistent with a geometric progression of
orbital sizes (constant $\Lambda$). Second, the observed multi-planet
systems exhibit regular orbital spacing \citep{weissa,weissb}, where
the values generally fall in the range $1.2<\Lambda<1.8$ (e.g.,
\citealt{rowe}; see also Figure 4 of \citealt{adams2019}). This
finding is consistent with theoretical expectations, which suggest
that forming planets will experience at least limited convergent
migration, but cannot become too close without becoming dynamically
unstable. The minimum separations for dynamical stability correspond
to $\Lambda\sim1.2-1.4$ for the {\it Kepler} sample, where the systems
are found to be stable \citep{puwu}.  Moreover, convergent migration
often leads to adjacent planets approaching mean motion resonance,
where the 3:2 and 2:1 commensurabilities are the most prevalent, and
correspond to $\Lambda\sim1.3$ and 1.6, respectively.  Although the
observed systems often have period ratios that are approximately 
given by ratios of small integers, they are generally not in resonance
\citep{fabrycky}. Finally, we note that the value of $\Lambda$ must
also depend on the planet masses, where the Hill stability boundary
provides a lower limit \citep{petit}.\footnote{Within the framework 
of this stability-sculpted picture, the uniformity of $\Lambda$ 
observed for the {\it Kepler} sample of multi-planet system arises 
in part because of the uniformity of the planetary masses. }

Note that we do not include the stellar spin as part of the energy
function (${\cal E}_{\rm spin}$ = $I\Omega_\ast^2/2$) or its
contribution to the angular momentum (${\cal L}_{\rm spin}$ =
$I\Omega_\ast$). This approach thus assumes that the planetary orbits
are sufficiently distant that coupling to the star, through tidal
effects or other mechanisms, is negligible over the time scales 
over which planetary masses are determined. 

Following previous treatments, we define dimensionless 
quantities according to   
\be
f = {m_1 \over \mtot}, \qquad 1-f = {m_2\over \mtot}, \qquad
{\rm and} \qquad a = {a_1 \over R_\ast} \,,
\label{fadefs} 
\ee
where $R_\ast$ is the stellar radius. The expression for the energy
then takes the form 
\be
{\cal E} = - {GM_\ast\mtot\over2R_\ast} 
\left\{ {1\over a} \left[f+{1-f\over\Lambda}\right]\right\} 
- \selfg {G \mtot^2 \over \rplan} 
\left\{ {\rplan \over R_1} f^2 + {\rplan \over R_2} (1-f)^2 \right\} \,,  
\ee
where $\rplan$ is a planetary radius (see below), and the 
angular momentum can be written 
\be
{\cal L} = \mtot (G M_\ast R_\ast)^{1/2} \sqrt{a} 
\left\{ f + (1-f) \sqrt{\Lambda} \right\} \,. 
\ee
Next we divide out the leading factors that define the energy and
angular momentum, thereby leaving behind the dimensionless expressions 
\be
E = {2 R_\ast {\cal E} \over GM_\ast \mtot} = 
- {1\over a} \left[f+{1-f\over\Lambda}\right]
- B \left[ f^2 + (1-f)^2 \right] \,, 
\label{edef} 
\ee
and 
\be
L = { {\cal L} \over \mtot (GM_\ast R_\ast)^{1/2}} = 
\sqrt{a} \left[ f + (1-f) \sqrt{\Lambda} \right] \,. 
\label{ldef} 
\ee
For simplicity, we have assumed that both of the forming planets have
the same radius, $R_1=\rplan=R_2$ (since the radius varies relatively
slowly with mass).\footnote{In the more physically realistic case, 
the planetary radius is expected to scale with planet mass. Since the
forming planets are not differentiated, the expected scaling is
$\rplan\sim m_j^{1/3}$. This complication would result in the
exponents of the second term of equation (\ref{edef}) being 5/3
instead of 2.} The dimensionless constant $B$ is defined as 
\be
B \equiv 2\selfg {R_\ast \over \rplan} {\mtot \over M_\ast} 
\approx 0.003 \left({\mtot\over10M_\oplus}\right)\,, 
\label{bdef} 
\ee
where we have assumed solar properties ($R_\ast=R_\odot$,
$M_\ast=M_\odot$) for the star to obtain the numerical value. 

\subsection{Extremum of the Energy} 
\label{sec:dfirst} 

The first step is to find the first variation of the energy. We can
write the semi-major axis in terms of the angular momentum through
equation (\ref{ldef}), so that the expression for the energy
(\ref{edef}) becomes 
\be
E = - \left[f+{1-f\over\Lambda}\right]
\left[ f + (1-f) \sqrt{\Lambda} \right]^2 
- BL^2 \left[ f^2 + (1-f)^2 \right] \,.
\label{enew} 
\ee 
Note that we have absorbed a factor of $L^2$ in the definition of
energy so that $EL^2\to E$. In this form, angular momentum is 
explicitly conserved, but its value only affects the relative 
size of the self-gravity term. 

We first take the derivative and set it equal to zero, 
\be
{dE \over df} = - BL^2 \left[4f - 2\right] 
- \left[1- {1\over\Lambda}\right]
\left[ f + (1-f) \sqrt{\Lambda} \right]^2  \qquad \qquad 
\label{quadstart} 
\ee
$$
\qquad \qquad 
- \left[f+{1-f\over\Lambda}\right]
2 \left[ f + (1-f) \sqrt{\Lambda} \right]
\left[ 1-\sqrt{\Lambda} \right] = 0\,. 
$$
The mass fraction is thus given by the solution to this quadratic
equation (in $f$). The coefficients depend on the orbital spacing
$\Lambda$ and the composite parameter $BL^2$ that determines the
relative importance of self-gravity (for a given angular momentum 
in the system). Because the coefficients are complicated functions 
of $(\Lambda,B)$, we use a collection of definitions for simplicity. 
Toward this end, we define 
\be
\Lambda \equiv x^2\,, \qquad \sqrt{\Lambda} \equiv x\,, \qquad 
{\rm and} \qquad 
\Upsilon \equiv 2 B L^2 {\Lambda \over (\sqrt{\Lambda}-1)^2} 
= {2 B L^2 x^2 \over (x-1)^2} \,.
\label{paramdef} 
\ee 
After some algebra, the equation (\ref{quadstart}) that 
determines the optimal mass fraction becomes 
\be
f^2 - f {2(2x^2+2x-1-\Upsilon) \over 3(x^2-1) } 
+ {(x^2+2x-\Upsilon) \over 3(x^2-1) } = 0 \,.  
\ee
This quadratic equation can then be written in the generic form 
\be
f^2 - b f + c = 0 \,, 
\label{generic} 
\ee
where the coefficients are given by  
\be
b \equiv {2(2x^2+2x-1-\Upsilon) \over 3(x^2-1) } 
\qquad {\rm and} \qquad 
c \equiv {(x^2+2x-\Upsilon) \over 3(x^2-1) } \,.  
\label{coeffdef} 
\ee
Note that we define $b$ such that the quadratic equation 
(\ref{generic}) has a minus sign. The roots are given by  
\be
f = {b \pm \sqrt{b^2 - 4c} \over 2} \,. 
\label{quadroots} 
\ee
We can show (see the following section) that the minus root
corresponds to an energy minimum (whereas the plus root is a
maximum). We can also show that in the limit $\Upsilon\to0$, 
where we ignore the self-gravity of the planets, the minus 
root reduces to the form found in previous work, i.e.,  
\be
f_{-} \to f_0 = {x^2 + x - 2 \over 3(x^2-1)} = 
{\Lambda + \sqrt{\Lambda} - 2 \over 3 (\Lambda-1)} \,.
\label{oldresult} 
\ee
Note that $f_0\to1/2$ in the limit $\Lambda\to1$, whereas $f_0\to1/3$
in the opposite limit $\Lambda\to\infty$ (note this latter result also
holds for $\Upsilon\ne0$). In the absence of self-gravity, the mass
fraction is thus a slowly varying function of the orbital separation
and is always relatively close to $f_0=1/2$ (equal planetary masses).
In the limit of wide separations $\Lambda\gg1$, the planets are
unlikely to interact enough to find their tidal equilibrium state, 
so this limit does not apply to the observed planetary systems. 

Notice also that the relevant root $f_{-}\to0$ when the
constant $c\to0$, which in turn corresponds to the condition
\be
x^2+2x = \Upsilon \,.   
\label{upcrit} 
\ee
From equations (\ref{paramdef}) and (\ref{upcrit}), it follows 
that this critical condition becomes 
\be
BL^2 = {(x+2)(x-1)^2 \over 2x} = 
{(\sqrt{\Lambda}+2)(\sqrt{\Lambda}-1)^2 \over 2\sqrt{\Lambda}} \,.
\label{bcritical} 
\ee
Using the definition of $B$ from equation (\ref{bdef}) and evaluating
the angular momentum for $f=0$ using equation (\ref{ldef}), the 
quantity $BL^2$ can be written in the form 
\be
BL^2 = \left( 2\selfg {\mtot \over M_\ast} {R_\ast \over \rplan} \right) 
\left( {a_2 \over R_\ast} \right) = 
2\selfg {\mtot \over M_\ast} {a_2 \over \rplan} \,. 
\label{bldef} 
\ee
Note that the composite parameter $BL^2$ is closely related to 
the Safronov number $\Theta\equiv(m_{\rm p} a)/(M_\ast \rplan)$, 
where the escape velocity replaces the velocity dispersion. 
The critical mass threshold $\mcrit$ thus takes the form  
\be 
\boxed{\quad
\mcrit =  M_\ast \left({\rplan\over a_2}\right) 
{(\sqrt{\Lambda}+2)(\sqrt{\Lambda}-1)^2 \over 4\selfg\sqrt{\Lambda}} 
\quad} 
\label{threshold} 
\ee
For typical values $\Lambda=1.6$, $M_\ast=1M_\odot$, $a_2$ = 0.1 AU,
$\rplan=3R_\oplus$, and $\selfg=0.5$, we find the critical value
$\mcrit\approx40M_\oplus$. 

Keep in mind that the threshold mass $\mcrit$ depends on the stellar
mass $M_\ast$, the spacing $\Lambda$, and especially the semimajor
axis $a_2$. For large semimajor axes, the threshold mass becomes
small, and planetary pairs tend to be supercritical. For masses well
above $\mcrit$, the binding energy of the planet dominates, and the
system can make the total energy maximally negative by putting all of
its mass into a single self-gravitating bucket. In this regime, energy 
optimization becomes nearly independent of orbits and depends almost 
entirely on the individual planets.
 
\begin{figure} 
%\figurenum{1} 
\centerline{ \includegraphics[width=0.85\textwidth]{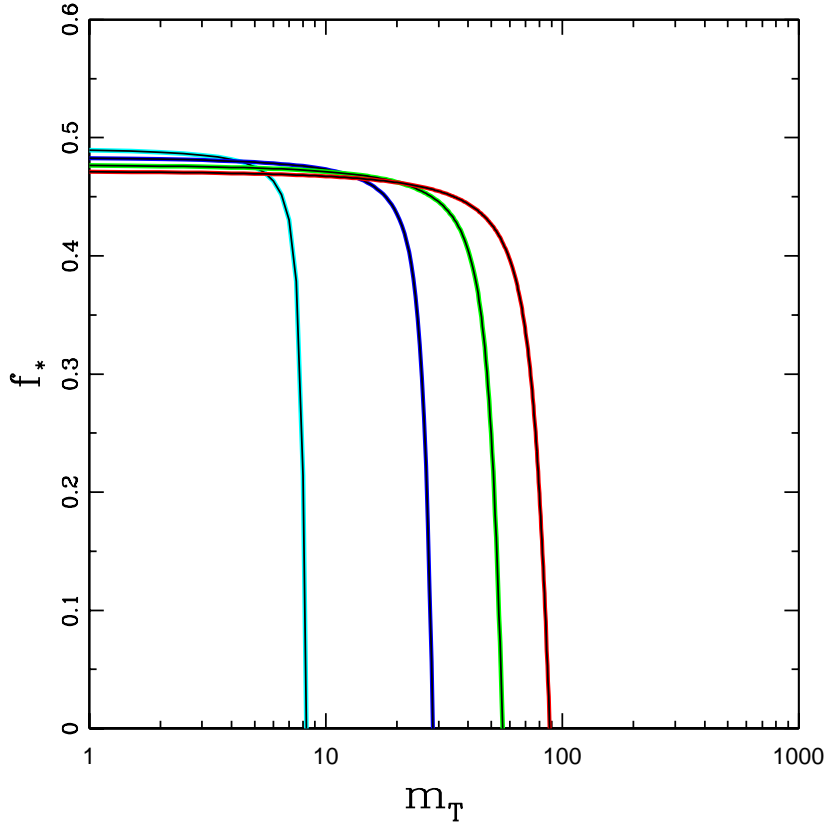} } 
\vskip-1.5truein
\caption{Optimized value of the mass fraction $f$ (of the inner 
planet) as a function of the total mass $\mtot$ of the planetary pair
(in units of $M_\oplus$).  For the sake of definiteness, the stellar
mass $M_\ast=1M_\odot$, the semimajor axis $a_2$ = 0.1 AU, and the
planet radius $\rplan$ = 3 $R_\oplus$. The four curves correspond to
different values of the orbital spacing parameter $\Lambda$ = 1.25 
(cyan), 1.5 (blue), 1.75 (green), and 2.0 (red), from left to right 
in the figure. Note the relatively sharp transition from $f\sim1/2$ 
to $f\to0$ as the total mass $\mtot$ increases. } 
\label{fig:fvb} 
\end{figure}  

Figure \ref{fig:fvb} shows the mass fraction $f$ as a function of the
total mass $\mtot$. Results are shown for four values of the orbital
spacing parameter $\Lambda$ = 1.25 -- 2, with the remaining parameters
the same as those used in the above estimate. The figure shows that
the transition from systems with nearly equal mass planets
($f\sim1/2$) to systems where one planet dominates the mass budget
($f\to0$) is relatively sharp. The mass threshold for this transition
increases with increasing orbital spacing, but is of order tens of
$M_\oplus$ for the expected (observed) values of the spacing parameter. 

\begin{figure} 
%\figurenum{2} 
\centerline{ \includegraphics[width=0.85\textwidth]{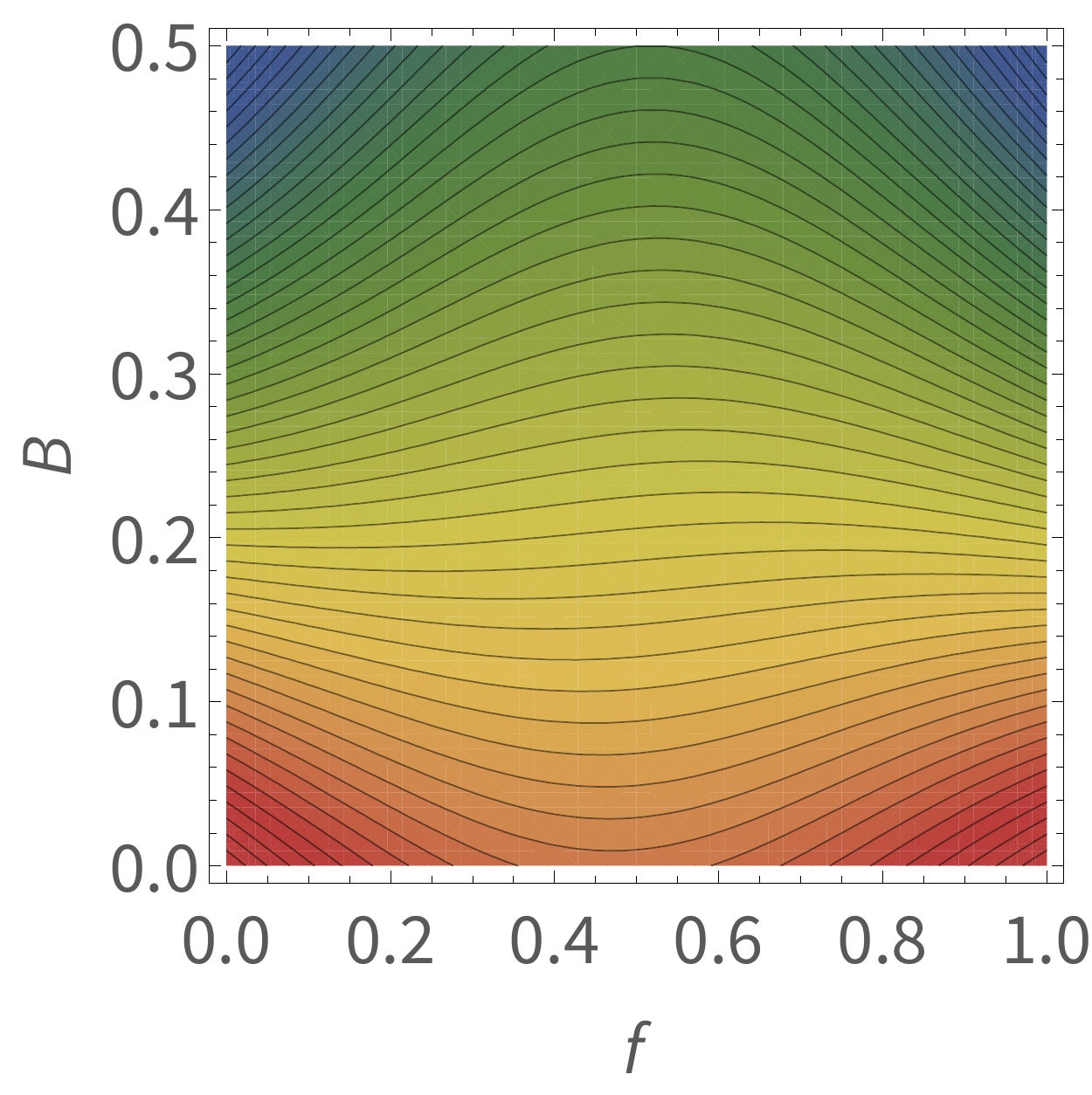} } 
%\vskip-0.25truein
\caption{Contour plot showing the energy as a function of the mass
fraction $f$ and the dimensionless parameter $B$ that specifies the
relative importance of self-gravity of the planets. The orbital
spacing parameter $\Lambda=2$.  Note that the contours of constant
energy are concave upward for small values of $B$, leading to
optimized values near $f\sim1/2$. In contrast, the contours are
concave downward for larger $B$, which favors $f\sim0$ and $f\sim1$. }
\label{fig:contour} 
\end{figure}  

Figure \ref{fig:contour} shows a contour plot of the energy as a
function of the mass fraction $f$ and the parameter $BL^2$ that
provides a measure of the importance of self-gravity in the system
(where we let $BL^2\to B$ to simplify the notation). For the sake of
definiteness, the orbital spacing parameter $\Lambda$ = 2 for this
figure. The contours of constant energy are concave upward for small
values of $B$ and transition to concave downward for larger values of
$B$. More specifically, the plot shows that we get a well defined
minimum at $f\sim1/2$ (equal planet masses) in the limit $B\to0$. For
larger $B\sim1$, the minimum becomes a well-defined maximum, and mass
fractions $f\to0,1$ are favored (with the true minimum moving to $f=0$
at the critical value of $B$ as given by equation
[\ref{bcritical}]). However, in the intermediate regime with
$B\sim1/4$, although the minimum value of the energy favors $f\to0$,
the energy difference is small, as depicted by the mostly horizontal
contours in the center of Figure \ref{fig:contour}. As a result, even
though the proclivity for equal mass planets disappears for
sufficiently large total masses $\mtot$ (equivalently $B$), the
preference for unequal mass planets remains weak until the total mass
$\mtot$ becomes significantly larger than the critical value $\mcrit$ 
(given by equation [\ref{threshold}]).

\subsection{Second Variation} 
\label{sec:dsecond} 

The analysis presented in the previous subsection specifies the 
equilibrium point of the system. In order to show that this critical
point corresponds to a minimum of the energy, we must consider the
second variation \citep{hesse,abrasteg}.  As shown above, the 
first derivative $dE/df$ is a quadratic function of the mass 
fraction $f$ and can be written in symbolic form
\be
{dE\over df} = - A \left[ f^2 - b f + c \right] = 0 \,, 
\ee
where we choose the sign of the linear term in the square brackets 
to be negative, and where the coefficients $b$ and $c$ are given 
by equation (\ref{coeffdef}). As written, the leading coefficient 
$A \equiv 3(x-1)^2 (x^2-1)/x^2 > 0$. The solution for the extremum 
is thus given by equation (\ref{quadroots}). The second derivative 
takes the form 
\be
{d^2E\over df^2} = - A \left[ 2 f - b \right] = 
- \pm A \left[b^2 - 4c\right]^{1/2} \,,
\ee
where the expression is evaluated at the critical point in the second
equality.  The negative root (of the quadratic whose solution
[\ref{quadroots}] specifies the extremum) thus leads to a positive
second derivative and hence an energy minimum, whereas the other root
corresponds to an energy maximum. As a result, the relevant root is
given by 
\be
f_\ast = {b - \left[b^2 - 4c\right]^{1/2} \over 2} \, .
\ee
Moreover, the mass fraction $f_\ast\to0$ when $c\to0$, which 
leads to the critical mass threshold of equation (\ref{threshold}). 

\section{Application to Planetary Systems} 
\label{sec:apply} 

The tidal equilibrium states found in the previous section are robust
in that they are independent of the dissipation processes that allow
planetary systems to evolve toward configurations of lower energy. On
the other hand, not all physical systems will achieve their optimal
states of minimum energy. In addition, the enforced constraints on the
optimization procedure (conservation of angular momentum, constant
total mass, and fixed orbital spacing) will not necessarily hold under
all circumstances. As a result, we examine the degree to which optimum
energy states can be attained by taking into account evolutionary
considerations (Section \ref{sec:evolve}). We then discuss the
conditions required for the underlying assumptions to be applicable
(Section \ref{sec:assume}).

In order for the minimum energy states of this paper to apply to
observed planetary systems, at least two conditions must hold: The
first is that a pair of planets can maintain a constant period ratio
$\Lambda$ while changing their radial distance from the star. It is
well established that planetary pairs often execute this type of
behavior while migrating in (or near) resonance (e.g., see
\citealt{peale76} for a physical explanation of resonant capture and
migration). The second requirement -- which is more novel -- is that
the forming planets can jointly apportion the available mass $\mtot$
between the two members. In other words, the planets must have some
type of agency that allows for the division of the material, where
this allocation is carried out so that energy is minimized. 

\subsection{Evolutionary Considerations} 
\label{sec:evolve} 

This section outlines an evolutionary scenario to illustrate how
forming planetary pairs could realize their minimum energy states as
they grow, subject to the two conditions presented above. Keep in
mind, however, that the optimized states found above are independent
of this or any evolutionary trajectory. Here we consider a forming
pair of planets, where both planets can grow by accreting mass from
the background circumstellar disk. We are thus implicitly assuming
that the planets are formed from the bottom up, through the
accumulation of mass, but otherwise invoke no specific mechanism.

At a given time, let the masses of the planets be given by $m_1$ and
$m_2$, where $m_1+m_2=\mtot$, and where the latter quantity is
considered to increase with time. Now suppose that an increment of
mass $\delta{m}$ is transferred from the disk to the planets. In order
to conserve the total angular momentum, before and after the accretion
of the new material $\delta{m}$, the following condition must be met 
\be 
\sqrt{a_i} \left[ m_{1i} + m_{2i} \sqrt{\Lambda} \right]
+ \sqrt{b} \,\,\delta{m} = 
\sqrt{a_f} \left[ m_{1f} + m_{2f} \sqrt{\Lambda} \right] \,,
\label{consang} 
\ee
where the subscripts $i(f)$ denote the initial (final) values, and
where $b$ is the semimajor axis of the mass increment before it is
accreted by the planets. To conserve total mass, the individual 
planet masses must obey the condition 
\be
m_{1i} + m_{2i} + \delta{m} = m_{1f} + m_{2f} \,. 
\ee
For any allowed final values of the planet masses, the semimajor axis
$a_f$ can always adjust to satisfy conservation of angular momentum
(\ref{consang}) while keeping the orbital spacing $\Lambda$ constant.
As a result, the planets change their radial locations at constant
$\Lambda$, thereby invoking the first requirement outlined above.

As mass is transferred from the disk to the forming planets, the total
mass of the planetary pair increases.  If the planets can reach their
minimum energy state as they grow, while simultaneously maintaining 
orbital period commensurability, then for a given mass $\mtot$, 
the individual planetary masses must be given by  
\be
m_1 = f(\mtot) \mtot \qquad {\rm and} \qquad 
m_2 = \left[ 1 - f (\mtot)\right] \mtot \,,
\label{massvtime} 
\ee
where the optimized mass fraction $f(\mtot)$ is a decreasing function
of the total mass (see Section \ref{sec:dfirst}).  

\begin{figure} 
\centerline{ \includegraphics[width=0.85\textwidth]{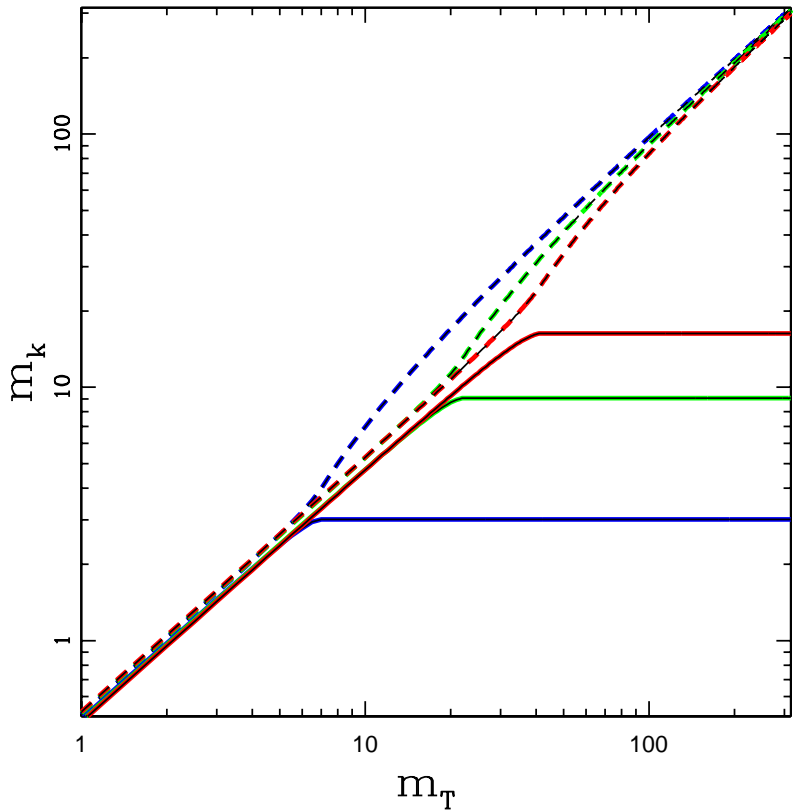} } 
\vskip-1.5truein
\caption{Masses $m_k$ of the two members of a planetary pair as a 
function of the total mass $\mtot$, which is considered here to be 
increasing with time.  All masses are given in $M_\oplus$.  The solid
curves show the mass of the inner (smaller) planet, whereas the dashed
curves show the mass of the outer (larger) planet. Results are shown
for three values of the orbital spacing parameter $\Lambda$ = 1.25
(blue, left), 1.5 (green, middle), and 1.75 (red, right). }
\label{fig:mevolve} 
\end{figure}

The discussion thus far assumes that equation (\ref{massvtime}) can be
satisfied at all stages of evolution.  At early times, when the total
mass is small $\mtot\ll\mcrit$, the mass fraction $f$ depends only on
$\Lambda$, and this balance can be achieved with both planets gaining
mass.  However, if the total mass exceeds the critical mass, then the
optimal mass fraction $f\to0$, which would imply that the mass of the
inner planet $m_1\to0$. But the smaller planet already has some mass
at this stage of evolution, so the system cannot achieve its optimal
energy state without removing mass from the planet. Instead of losing
mass, the inner planet is likely to maintain a constant mass, as
determined by its value at this transition point, but stop growing.
The other planet then accretes essentially all of the additional mass
from that epoch onward. This scenario is illustrated in Figure
\ref{fig:mevolve}, which shows the masses of both planets as function
of $\mtot$, where the mass of the inner planet reaches a constant
value.

In the scenario depicted by Figure \ref{fig:mevolve}, the mass of the
inner planet reaches an asymptotic value $m_1\to m_\infty$, which we
estimate as follows.  The optimum mass of the inner planet $m_1$ 
(from equation [\ref{massvtime}]) can be considered as a function of 
$\mtot$.  The function increases for small values of $\mtot$, reaches
a maximum, and then decreases for sufficiently large $\mtot$ (since
$f\to0$ in the limit $\mtot\to\mcrit$). The point where the inner
planet reaches its final mass $m_\infty$ is thus given by the
condition 
\be
{d m_1 \over d\mtot} = f(\mtot) + \mtot {df \over d\mtot} = 0\,.
\label{maxcon} 
\ee
For small values of $\mtot$, the mass fraction $f$ is nearly constant,
so that the derivative $dm_1/d\mtot$ is positive.  As the total mass
approaches the critical value $\mtot\to\mcrit$ (see equation
[\ref{threshold}]), the quantity $df/d\mtot$ becomes large and
negative (see Figure \ref{fig:fvm}) while $f$ itself becomes small, 
so that $dm_1/d\mtot$ is negative at $\mtot=\mcrit$. In practice, the
condition of equation (\ref{maxcon}) is realized for values of $\mtot$
comparable to but somewhat smaller than $\mcrit$.  One can find a
closed form solution for the value of $\mtot$ at this transition
point, but it requires the solution to a cubic equation and the
resulting expression is awkward (note that Figure \ref{fig:mevolve}
uses the numerical solution). A reasonable approximation can be found
by expanding the equations of Section \ref{sec:dfirst} to leading
order to obtain the expression
\be
m_\infty \approx {1 \over 2} \mcrit \, 
\left[ 1 - {1 \over 2} \left( 
{\Lambda - 1 \over \Lambda + 2 \sqrt{\Lambda} }
\right)^{1/2} \right]^2 \,,
\label{asympmass} 
\ee
where the critical mass is given by equation (\ref{threshold}). 
This expression for the asymptotic value $m_\infty$ for the mass 
of the inner planet agrees with the full solution to within 
$\sim8\%$ for typical planetary systems (those depicted in 
Figure \ref{fig:mevolve}). Note that for late times, the mass 
fraction in this scenario takes the form $f \to m_\infty/\mtot$. 

In this problem, the quantity $BL^2$ represents a control parameter.
The critical value from equation (\ref{bcritical}) is a bifurcation
point \citep{strogatz}, where the preferred state transitions from
nearly equal mass planets to one planet dominating the mass supply.
However, planetary systems will not always be able to reach their
optimal configurations.  Consider a system with slowly increasing
total mass $\mtot$.  At a given value of $\mtot$, the planets find
their optimal configuration, which will have nearly equal masses at
early times. In a perfect system, but one with evolutionary
constraints, the mass of the smaller planet will stop growing at
$m_\infty$. In addition, the dynamics of the formation processes
(although unspecified in this treatment) will involve some type of
time evolution -- a flow -- for the mass fraction $f(t)$.  As the
system approaches the bifurcation point ($\mtot\to\mcrit$) from below,
this flow is pushing the system toward the equal mass state. After the
critical point is reached, this flow continues at first, but it
eventually changes after the system becomes sufficiently
supercritical. This type of residual behavior, sometimes known as the
`ghost in the bifurcation' \citep{strogatz}, prevents the system from
fully realizing its minimal energy configuration, and results in
scatter in the resulting system properties.

\subsection{Assessment of Assumptions} 
\label{sec:assume}

This optimization scenario of this paper assumes that the forming
planets conserve angular momentum and maintain fixed orbital spacing.
The calculation also makes implicit assumptions about the relevant
time scales. This section examines these assumptions and discusses 
the conditions necessary for planetary systems to realize these 
minimum energy states. 

In order for the planetary pair to maintain its optimal energy
configuration as the bodies grow, the energy dissipation time scale
must be much shorter than the time scale for planet formation.
Moreover, the two planets must be able to coordinate their mass
intake.  The natural dynamical time of the system is the orbit time,
and any gas in the system communicates through pressure perturbations,
which travel at the sound speed.  A successful realization of this
optimization scheme thus requires the ordering of time scales 
\be
t_{\rm sc} \sim t_{\rm orb} \ll t_{\rm diss} \ll t_{\rm form} \,,
\label{torder} 
\ee
for the sound crossing, orbital, dissipation, and formation time
scales, respectively. The orbit time and sound crossing time are
typically a fraction of a year, whereas formation time scales are of
order millions of years. As a result, the dissipation time can vary
over a relatively wide range and still satisfy the intermediate 
asymptotic requirement of equation (\ref{torder}).
 
The optimization procedure of this paper assumes that the planetary
pair conserves angular momentum and total mass, while keeping the
orbital spacing fixed. Although angular momentum of the entire system
is conserved, planet formation takes place within a circumstellar disk
containing the planets themselves, along with rocks, dust, and gas.
Since the planets under consideration are often super-Earths, they are
primarily made of rocky material.  More specifically, recent
observations \citep{fulton} indicate that the physical structure of
super-Earths is consistent with rocky cores surrounded by low-mass
envelopes of hydrogen and helium. As a result, it remains possible for
angular momentum of the forming planets to be transferred to the
gaseous component, which is eventually removed from the system. The
resulting decrease in angular momentum of the planets can lead to
migration, which, when paired with planet-planet interactions, can
maintain fixed orbital spacing. 

In the limit of small total mass $\mtot<\mcrit$, the optimization
procedure is independent of the angular momentum (equation
[\ref{enew}]), so that $f\approx f_0\sim1/2$ (equation
[\ref{oldresult}]). Decreasing angular momentum leads to an increasing
critical mass scale (equation [\ref{threshold}]). We thus have the
following possible scenarios: If planetary migration is limited, so
that $(\Delta a)/a\ll1$ and hence $(\Delta L)/L\ll1$, then angular
momentum is essentially conserved, any small changes can help maintain
the fixed orbital spacing, and the approximations of the previous
section are valid.  In the case where the planets migrate over longer
distances, with larger $(\Delta L)/L$, the critical mass scale
$\mcrit$ increases as the planets move inward.  Since the planets
start with small masses, where $\mtot<\mcrit$ and $f\approx
f_0\sim1/2$, this decreasing angular momentum (increasing critical
mass) allows the planets to maintain their well-ordered state (with $f
\sim f_0\sim1/2$) over longer times. In this case, the relevant
critical mass scale is given by equation (\ref{threshold}) evaluated
at the final value of $L$ (equivalently $a$). In addition to the
ordering of time scales given by equation (\ref{torder}), this
scenario requires energy dissipation to occur more rapidly than
planetary migration, so that $t_{\rm diss}\ll$ 
$t_{\rm mig}$.\footnote{As one example, if the dissipation is 
time scale is limited by that of wave propagation, then the 
ratio $t_{\rm diss}/t_{\rm mig}\sim(h/a)^2\ll1$, where $h$ is 
the disk scale height \citep{tanakaward}. }

One can also consider the case where planet formation takes place
within a gas-free environment, corresponding to later evolutionary
stages. If planetesimals can scatter off of the growing planets, they
can be ejected or accreted by the central star, so that the total
angular momentum can change in principle. In practice, the planets of
interest reside in the inner part of their solar systems, where
ejection is suppressed because the required speed is much larger than
the escape speed from the planetary surfaces.  Ejection will be
important for planets forming on sufficiently distant orbits, roughly
given by
\be
a \gta {M_\ast \over m_p} R_p \sim 2\,\,{\rm AU}\,. 
\ee
Even in the inner solar system, the validity of the optimization
procedure requires that the forming planets accrete rocky material
with high efficiency, i.e., with little loss from ejection or 
accretion by the star.

In a gas-free environment, migration can no longer lock planets into
mean motion resonance, so that the orbital spacing $\Lambda$ could
vary. Numerical simulations show that planets growing within a
planetesimal disk will roughly maintain their orbital spacing if they
are sufficiently distant, but will experience orbital divergence if
the orbits are too close (e.g., \citealt{kokida95}).  The boundary
between these two types of behavior corresponds to separations of
$\sim5$ Hill radii, and results in typical separations of $\sim10$
Hill radii \citep{kokida98}, which is comparable to the observed
orbital spacing \citep{rowe}. The results of the optimization
procedure vary slowly with changes in the orbital spacing. In the 
low mass limit, the optimal mass fraction $f_0\to1/2$ in the limit 
$\Lambda\to1$ and $f_0\to1/3$ as $\Lambda\to\infty$. The critical mass
scale $\mcrit\propto\Lambda$ in the limit of large $\Lambda$.  As
result, if planets forming in a gas-free environment increase their
orbital spacing due to interactions with planetesimals, the optimum
mass fraction will slowly decrease, but the critical mass scale 
$\mcrit$ will also increase so that the system tends to remain in the
low-mass limit with mass fraction $f \sim f_0$. In addition, this 
scenario allows $\mcrit$ to plausibly become larger than the local 
mass budget of solid material. 

For completeness, we consider the limiting cases where the constraints
are removed entirely. For systems in which angular momentum is not
conserved, the lowest energy state is given by the minimum of equation
(\ref{edef}), where both the mass fraction $f$ and semimajor axis $a$
can vary.  The energy is unbounded from below: The system can always
evolve to a lower energy state by moving the planets inward and hence
decreasing $a$. This scenario results in all planets being accreted by
the central star. In the other case, where the constraint of fixed
orbital spacing is removed, the lowest energy state is given by the
minimum of equation (\ref{enew}), where the mass fraction $f$ and the
spacing $\Lambda$ can vary. In the low mass limit $\mtot\ll\mcrit$,
this minimization procedure implies that $\Lambda\to1$, i.e., the
planets are predicted to merge. The complete removal of the
constraints thus leads to an optimal configuration with either no
planets (no angular momentum constraint) or a single planet (no
spacing constraint). 

\section{Comparison to Observations} 
\label{sec:obs} 

The analysis carried out in the previous section suggests that when
the total mass $\mtot$ of planetary pairs becomes sufficiently large,
the preferred value of the mass fraction $f$ switches from nearly
equal mass planets with $f\sim1/2$ to unequal mass planets with
$f\to0$. Previous work has already indicated that the mass uniformity
of planetary systems is compromised when the system contains a Jovian
planet \citep{songhu}, consistent with these results for systems with
large $\mtot$. For these high mass cases, since $f\to0$, the outer
planet becomes the larger member of the pair. This finding is also
consistent with results for planetary pairs in {\it Kepler}
multi-planet systems \citep{ciardi}, where the outer planet is more
likely to be larger if one or both planets is larger than Neptune. In
this section, we look for related observational signatures using the
observed mass fractions of planetary pairs.

The observational data used in this section are the same as in
\cite{adams2019}.  Briefly, the data were extracted from the publicly
available exoplanet
database\footnote{https://exoplanetarchive.ipac.caltech.edu}, which
includes 219 planetary systems with $\nump\ge3$ planets, where these
systems contain a total of 777 planets.  All of the 557 adjacent
planetary pairs found in the observational sample are used in this
simple analysis (see also \citealt{fabrycky,fangmargot,millholland,
  petit,puwu,rowe,tredong,songhu,weissa,weissb}).  Systems with only
two planets are not included here because they tend to have larger
spacing and mutual inclinations, and hence to not display the
regularity found in systems with larger $\nump$.  If both planet mass
and planet radius are not reported, then the mass is estimated from
the observed radius, or the radius is estimated from the measured mass
(see \citealt{adams2019} for further detail). One should keep in mind
that the mass-radius relation for exoplanets is a multi-valued
function \citep{wolfgang}, so that this approach is valid only in a
statistical sense.

\begin{figure} 
%\figurenum{3} 
\centerline{ \includegraphics[width=0.85\textwidth]{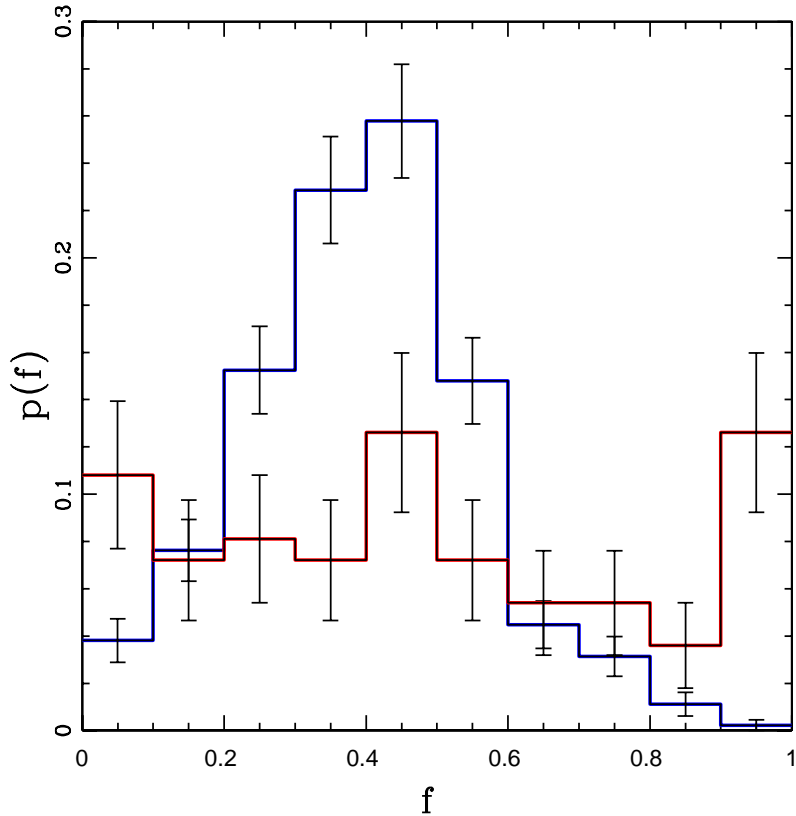} } 
\vskip-1.5truein
\caption{Distribution of mass fraction $f$ for the observed 
sample of planetary pairs found in extrasolar planetary systems 
with $\nump\ge3$ members. The blue histogram shows the normalized
distribution for pairs of low mass planets with total mass 
$\mtot=m_1+m_2<40M_\oplus$. The red histogram shows the 
corresponding distribution for higher mass planets with 
$\mtot=m_1+m_2>40M_\oplus$. The error bars represent the
$\sqrt{N}$-errors. The two distributions are clearly distinct. }
\label{fig:fdist} 
\end{figure}

For all of the planetary pairs in the sample, Figure \ref{fig:fdist}
presents two distributions of the mass fraction $f$. The two
histograms in the figure correspond to the low mass and high mass
parts of the sample, with $\mtot<40M_\oplus$ and $\mtot>40M_\oplus$,
respectively. The two distributions are manifestly different. The
figure includes the $\sqrt{N}$-errors, marked by error bars, which are
smaller than the differences between the two distributions. The
distribution for low mass planets shows a significant peak near
$f\sim1/2$, consistent with predictions for systems where the
self-gravity of planets is subdominant (equation [\ref{oldresult}];
see also \citealt{millholland,weissa}). The distribution for high mass
planets is nearly flat (within the estimated uncertainties), with
slight preferences for the largest and smallest bins (the most unequal
mass planets). This latter behavior is expected for systems with
sufficiently high total mass $\mtot$ (equation [\ref{threshold}]; see
also \citealt{songhu}).

For completeness, we note that the distributions shown in Figure
\ref{fig:fdist} include planetary pairs of all semimajor axes.
However, both the available mass in the original circumstellar disk
and the tendency for planetary pairs to experience runaway growth are
expected to increase with $a$. Here, the members of the high mass
sample span a wide range of semimajor axis, namely 0.15 AU $\le{a}\le$
12 AU. Nonetheless, the mean value of $a$ is larger for the high mass
sample ($\langle{a}\rangle\approx0.71$ AU) than for the low mass
sample ($\langle{a}\rangle\approx0.13$ AU).

\begin{figure} 
%\figurenum{4} 
\centerline{ \includegraphics[width=0.85\textwidth]{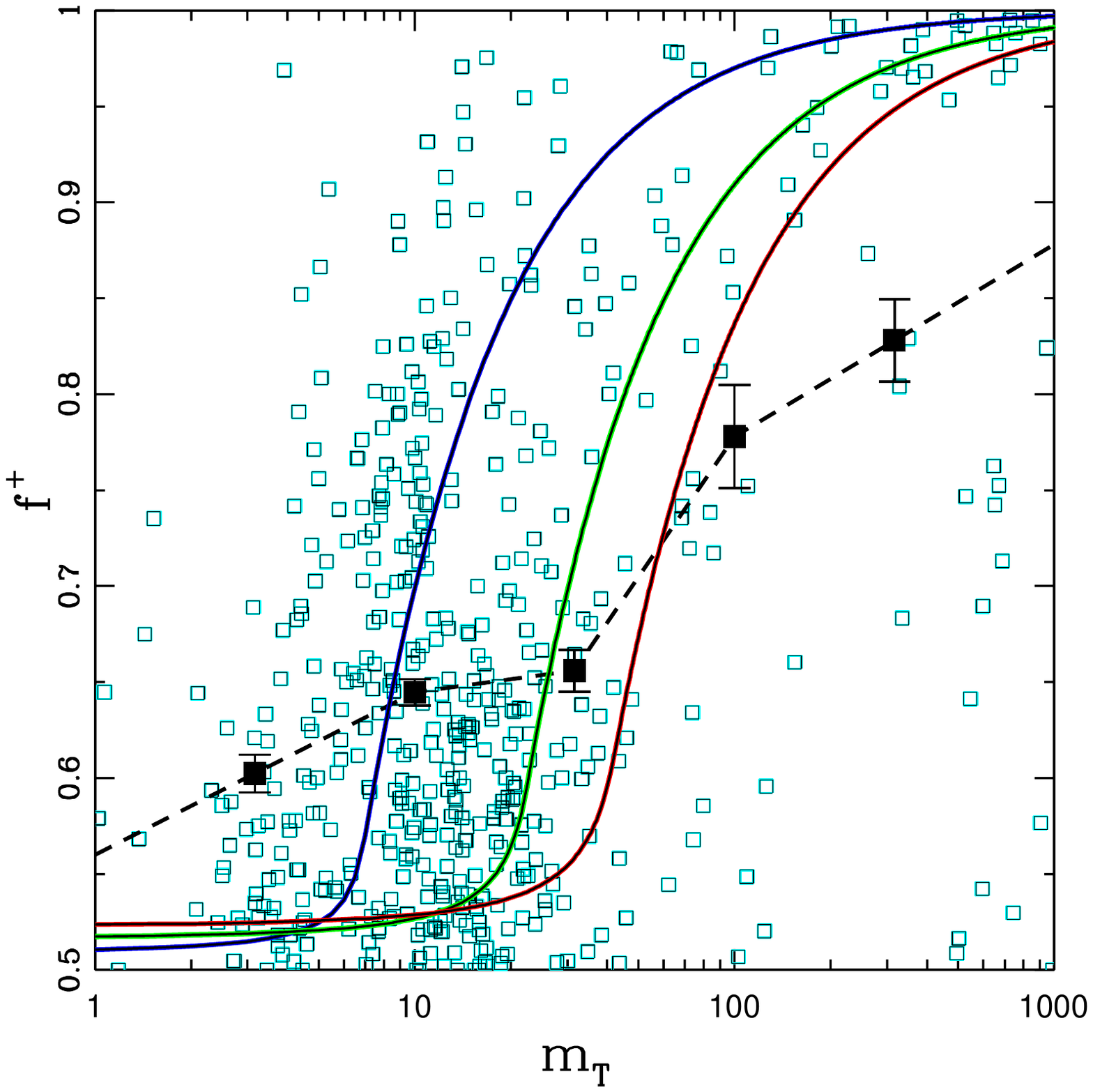} }
\vskip-1.5truein
\caption{Mass fraction $f^+$ of largest planet as a function 
of total mass $\mtot$ of the planetary pair. The alternative mass 
fraction $f^+$ is defined to be $f^+=f$ when the inner planet has 
the larger mass and $f^+=1-f$ when the outer planet is heavier. 
The open cyan squares show the mass fractions for all of the planetary
pairs in the observational sample. The solid black squares with error
bars show the binned version of the same data. The line dashed
connecting the binned data points has been extended to lower and
higher masses. The solid curves show the theoretically expected mass
fraction $f^{+}$, based on the evolutionary scenario of the previous
section, for three values of the orbital spacing parameter $\Lambda$ 
= 1.25 (left, blue), 1.5 (middle, green), and 1.75 (right, red).
Although the data show significant scatter, the mass fraction $f^{+}$
is an increasing function of the total planetary mass $\mtot$,
especially for $\mtot\gta40M_\oplus$, roughly consistent with the
expectations of this paper. } 
\label{fig:fvm} 
\end{figure}

The considerations of this paper suggest that planet masses should be
comparable for low-mass planetary pairs, but the mass ratios should be
much larger for sufficiently high total masses $\mtot$. Another way to
test this prediction is to plot the observed mass fractions as a
function of total mass $\mtot$ in the planetary pair, as shown in
Figure \ref{fig:fvm}. Since we want to include departures from the 
case of equal mass planets when either planet is larger, we define 
an alternative mass fraction $f^{+}$ according to 
\be
f^{+} \equiv {\rm Max} \left\{ f, 1-f \right\} \,. 
\ee
As defined here, this mass fraction must fall in the range 
$1/2\le f^{+} \le 1$. 

Figure \ref{fig:fvm} shows the expected trend that the mass fraction
$f^{+}$ for observed planetary pairs increases with total mass
$\mtot$.  For comparison, the theoretically expected mass fraction
$f^{+}$ is shown for three values of the (fixed) orbital spacing,
$\Lambda$ = 1.25, 1.5, and 1.75, from left to right in the
diagram. These curves are calculated from the evolutionary scenario of
Section \ref{sec:evolve}, where the inner/smaller planet grows as
long as it is energetically favorable, but does not decrease its mass.
Note that the observed data, depicted as open cyan squares, show a
significant amount of scatter. The figure also includes the binned
data, shown by the solid black squares with error bars. The binned
data indicate that the mass fraction $f^{+}$ is an increasing function
of $\mtot$.  Moreover, the mass fraction increases slowly for small
$\mtot$, but starts to increase more rapidly for larger total mass
$\mtot\gta40M_\oplus$. The observed threshold for mass fraction
increases is thus roughly consistent with that of equation
(\ref{threshold}). Note that the exact shape of the curve shown in
Figure \ref{fig:fvm} depends on the choice of bin size, although the
trend of $f^{+}$ increasing with total mass $\mtot$ is robust.
Moreover, the (binned) observational curve lies above the theoretical
values for low masses $\mtot\lta3M_\oplus$ and below for high masses 
$\mtot\gta300M_\oplus$. This discrepancy indicates that the forming 
planetary systems do not always reach their optimal energy 
configurations. 

If the masses of the planetary pairs were sampled randomly from a
known distribution, then the expected alternative mass fraction
$f^{+}$ could be determined and compared with those observed (as shown
in Figure \ref{fig:fvm}). Although the underlying mass distribution of
the planets is not known, we can illustrate this procedure with a
representative example. Suppose that the planetary mass distribution
has the form 
\be
{dN \over dm} \propto {1 \over m} \qquad {\rm for} \qquad 
m_0 \le m \le \mtot \,,
\ee
where $m_0$ is the lower-mass cutoff. The actual planetary mass
distribution is undoubtedly more complicated, but is subject to a host
of selection biases (e.g., \citealt{massdist}), so its form remains
under study. For this distribution, the expectation value of the
alternative mass fraction is then given by 
\be
\langle f^{+} \rangle =  
{\ln (\mtot/2m_0) + m_0/\mtot \over \ln (\mtot/m_0) } \,.
\ee
If we choose $m_0=1M_\oplus$, under the assumption that smaller
planets are not detected in the present sample, then we find that
$\langle f^{+} \rangle \approx 0.7$ for $\mtot\sim10M_\oplus$ and
increases to $\langle f^{+} \rangle \approx 0.9$ for
$\mtot\sim100-1000M_\oplus$. The expectation value of the mass
fraction is thus significantly larger than that observed for low-mass
planetary pairs. In contrast, for high-mass pairs, the observed 
mass fraction approaches the expectation value. As a result, the 
low-mass pairs show non-random (correlated) masses, consistent 
with the observational findings that motivated this work. On the 
other hand, this correlation is compromised for higher total 
masses, consistent with the theoretical results derived in 
Section \ref{sec:model}. 

The moons of the giant planets in our solar system provide another
setting for testing the predictions of this energy optimization
scenario. The four Galilean moons of Jupiter have nearly uniform
masses of $m$ = 0.015, 0.008, 0.025, and 0.018 $M_\oplus$, a spread of
orbital inclination angles $\Delta{i}<0.5^\circ$, and spacing
parameters $\Lambda$ = 1.6, 1.6, and 1.8 (from Io on the inside to
Callisto on the outside). Moreover, the critical mass predicted from
equation (\ref{threshold}) is approximately $\mcrit\approx0.06-0.07$
$M_\oplus$, comfortably larger than the masses of the moons. The
Galilean satellite system can thus be understood as a pair-wise
equilibrium state. In contrast, the moons of Saturn have non-uniform
masses and orbital spacing. The largest body --- Titan --- dominates
the satellite system with mass $m$ = 0.023 $M_\oplus$, which is
comparable to the critical mass estimated from equation
(\ref{threshold}), specifically $\mcrit$ = 0.024 (0.041) $M_\oplus$
for the pairing of Titan with Iapetus (Rhea). It is important to
stress that the mechanisms that lead to satellite formation are
complex; nonetheless, the observed properties are roughly consistent
with the optimization scenario considered here.

\section{Conclusion} 
\label{sec:conclude} 

\subsection{Summary of Results} 
\label{sec:summary} 

This paper determines the lowest energy states available to forming
planetary pairs subject to conservation of angular momentum, constant
total mass, and fixed orbital spacing. This work generalizes previous
treatments by including the self-gravity of the planets in the energy
budget (see Section \ref{sec:model}).  In the limit where the total
mass $\mtot$ contained in planets is small, the mass fraction of
planetary pairs approaches $f\approx1/2$ (see equation
[\ref{oldresult}]), corresponding to nearly equal mass bodies. As
shown previously, this configuration corresponds to circular orbits
confined to a single plane, and can be extended to systems with larger
numbers of planets (which are then predicted to have nearly equal
mass). As the total mass $\mtot$ increases, the mass fractions become
increasingly asymmetric, and formally $f\to0$ when the mass $\mtot$
exceeds a critical threshold $\mtot\gta\mcrit\sim40M_\oplus$ (given by
equation [\ref{threshold}]).  For pairs with larger available mass
$\mtot$, one planet can thus experience runaway growth, which results
in asymmetric masses, and allows for the formation of Jovian planets.

The planetary pairs found in currently observed multi-planet systems
have properties that are roughly consistent with the predictions of
this energy optimization scenario (see Section \ref{sec:obs}).
Specifically, the masses of members of planetary pairs become more
asymmetric as the total mass $\mtot$ increases (Figure \ref{fig:fvm}).
Similarly, the distribution of mass fractions $f$ shows a clear peak
near $f\sim1/2$ for low-mass pairs with $\mtot<40M_\oplus$; in
contrast, the distribution is nearly flat (uniform) for higher masses
$\mtot>40M_\oplus$ (Figure \ref{fig:fdist}). These observational
signatures are consistent with previous work that motivated this
analysis, where many systems show nearly equal mass planets on
regularly spaced orbits (e.g., \citealt{weissa}), whereas systems with
large planets break with this relationship \citep{songhu}. 

The current data show a significant amount of scatter, and only
$\sim200$ multi-planet systems (with $\nump\ge3$) have been detected,
so that more observations are necessary to explore these trends in
greater detail. Fortunately, new multiple planet systems --- and new
members --- are being discovered and characterized on a regular basis
(e.g., see \citealt{badenas,feng,hidalgo,nielsen,rodriguez}). Such
data can be used to update and generalize the correlations depicted in
Figure \ref{fig:fvm}. For example, it will be useful to plot the
generalized mass fraction $f^{+}$ as a function of the total mass
$\mtot$ of planetary pairs for various sub-populations as the data
become available.

The theory presented in this paper implies several predictions that
can be tested in the near-term as the number of detected systems
increases, and as the extant system parameters are improved. Such
advances will primarily occur through improved planetary mass
measurements, but also through detection (via transit timing) of
non-transiting planets that occupy apparent gaps in the known
multiple-planet systems (see \citealt{weissa,weissb} for some
examples). With more and better data, the dependence of $f^{+}$ on
$\Lambda$ can be tested, as well as the linear dependence of $\mcrit$
on $M_\ast$ (see equation [\ref{threshold}]).

\subsection{Discussion}
\label{sec:discuss} 

The pair-wise optimization scheme developed here (see also
\citealt{adams2019}) has important implications for our understanding
of extrasolar planetary systems.  The lowest energy states available
to low-mass planetary systems correspond to nearly equal mass planets
on circular orbits, all confined to the same orbital plane. For
planetary systems with larger masses, we have shown that it becomes
energetically favorable for one planet to dominate the mass supply. It
must be emphasized that these results follow from the basic principles
of energy minimization, conservation of angular momentum, and
conservation of mass (subject to the constraint of fixed orbital
spacing). Significantly, these results are independent of the any
specific dissipation mechanism and any particular paradigm invoked to
explain planet formation, such as Type I migration, pebble accretion,
streaming instability, and so on.

This optimization scheme not only explains the so-called peas-in-a-pod
properties of the compact multi-planet systems observed by the 
{\it Kepler} mission, but also applies over a wide range of mass
scales. The aforementioned systems typically have stellar host masses
$M_\ast\approx0.5-1M_\odot$ and (nearly) equal mass planets with
$\mplan\sim10M_\oplus$. In contrast, the seven planet system
associated with Trappist-1 \citep{gillon} displays similar uniformity
but its mass scales are smaller by an order of magnitude. More
specifically, the primary mass $M_\ast=0.08M_\odot$ and the chain of
nearly equal mass planets have $\mplan\approx1M_\oplus$. As outlined
in Section \ref{sec:obs}, the satellite systems of the giant planets
\citep{galileo} also display this phenomenon, but the masses are
smaller by factors of $\sim1000$: For example, note that Jupiter has
mass $m_J=10^{-3}M_\odot$ and the Galilean moons have masses of order
$m_{\rm gal}\sim0.01M_\oplus$ (equivalently, $10^{-3}\times
10M_\oplus$).  This phenomenon of regularly spaced, equal mass
orbiting bodies thus displays an intriguing degree of universality.

We also note that the giant planets in our solar system exhibit some
degree of regularity. Although their total masses vary appreciably,
all four planets are thought to have rocky cores with masses 
$m_{\rm cor}\sim10M_\oplus$ and these bodies exhibit relatively
uniform spacing with $\Lambda\approx$ 2, 2, and 3/2 for the three
adjacent pairs (using current orbits). According to many models for
the early evolution of the solar system \citep{tsiganis}, the giant
planets could have formed in a more compact configuration, with
tighter orbital spacing $\Lambda\sim3/2$. However, the critical mass
$\mcrit$ (from equation [\ref{threshold}]) decreases with semimajor 
axis, so that the cores of our giant planets are well above the
threshold and probably do not represent a pair-wise equilibrium state.

The concept of energy optimization, with varying individual masses
subject to constant total mass, can be applied in other contexts. One
interesting setting is that of binary stars. In one classic
application \citep{counselman,hut1980}, energy optimization of binary
systems with fixed mass leads to tidal equilibrium states where the
rotation rates of both stars are synchronous with the orbital angular
velocity, the orbit is circular, and the angular momentum vectors
point in the same direction. This problem can be generalized to find
the optimal mass fraction for the binary system, analogous to the case
considered here for planets. This calculation is beyond the scope of
this paper and will be presented elsewhere \citep{darwinbinary}.

The basic strategy of this paper is to identify underlying principles
--- here energy optimization --- that specify or constrain the
properties of planetary systems. We note that alternate approaches are
being developed. As one example, the orbit distribution of observed
systems can be described by a statistical model that assumes the
systems uniformly sample the stable portion of phase space
\citep{tremaine15}. The results of this paper could be used in such a
formalism: The `temperature' of the statistical model could be defined
in terms of the energy difference between the observed planetary pairs
and their minimum energy states (found here). 

The results of this paper have a number of additional implications.
Given the form of equation (\ref{threshold}) for the critical mass,
the pair-wise interaction effects considered in this paper favor equal
mass planets when they orbit close to their host stars. If the
observed {\it Kepler} planets, with uniform masses and spacing, had
formed in the outer parts of their solar systems and migrated inward,
then self-gravity effects could more easily disrupt the pair-wise
equilibrium state.  For $a\gta10$ AU, for example, the threshold mass
falls to $\mcrit\lta1M_\oplus$, which reflects the fact that
gravitational binding energy becomes a progressively more important
part of the energy budget with increasing distance from the star. As
a result, if this energy optimization scheme provides the explanation
for the uniformity of observed compact systems, then {\it in situ} 
formation is favored. A scenario where the range of migration is
limited can provide the explanation for the uniform orbital spacing by
moving planets toward mean motion resonances, but migration over large
distances (see also Section \ref{sec:assume}) is disfavored.

Given that the pair-wise equilibrium states should be less pronounced
for outer planets ($\mcrit$ decreases with increasing $a$), this
scenario makes another prediction: The mass fractions $f$ of planetary
pairs should favor $f\sim1/2$ for close orbits and should depart from
this uniform state for pairs with increasing semimajor axis $a$. In
other words, the peas-in-a-pod phenomenon observed for compact
planetary systems should not be prevalent for planets found in larger
orbits. This prediction should become testable as the observational 
sample expands to include planets with longer periods. 

\bigskip 
\textbf{Acknowledgments:} We would like to thank Juliette Becker,
Darryl Seligman, Chris Spalding, and Lauren Weiss for useful
discussions. We also thank an anonymous referee for constructive input
that improved the paper.  This work was supported through the
University of Michigan, the National Science Foundation (DMS-1613819),
the Air Force Office of Scientific Research (FA 0550-18-0028), NASA
(NNX16AB47G), the David and Lucile Packard Foundation, and the Alfred
P. Sloan Foundation.


\begin{thebibliography}{99}

\bibitem[Abramowitz \& Stegun(1972)]{abrasteg} 
Abramowitz, M., \& Stegun, I. A. 1972, Handbook of Mathematical
Functions (New York: Dover)

\bibitem[Adams(2019)]{adams2019}
Adams, F. C. 2019, MNRAS, 488, 1446 

\bibitem[Adams et al.(2020)]{darwinbinary} 
Adams, F. C., Batygin, K., \& Bloch, A. M. 2020, 
submitted to MNRAS 

\bibitem[Adams \& Bloch(2016)]{ab2016}
Adams, F. C., \& Bloch, A. M. 2016, MNRAS, 462, 2527

\bibitem[Adams \& Bloch(2015)]{ab2015}
Adams, F. C., \& Bloch, A. M. 2015, MNRAS, 446, 3676

\bibitem[Badenas-Agusti et al.(2020)]{badenas} 
Badenas-Agusti, M., G{\"u}nther, M. N., Daylan, T., et al. 2020
arXiv:2002.03958

\bibitem[Batalha et al.(2011)]{batalha} 
Batalha, N. M., Borucki, W. J., Bryson, S. T., et al. 2011, ApJ, 729, 27

\bibitem[Borucki et al.(2010)]{borucki}
Borucki, W. J., Koch, D., Basri, G., et al. 2010, Sci, 327, 977

\bibitem[Chiang \& Laughlin(2013)]{chiang} 
Chiang, E., \& Laughlin, G. 2013, MNRAS, 431, 3444 

\bibitem[Ciardi et al.(2013)]{ciardi} 
Ciardi, D. R., Fabrycky, D. C., Ford, E. B., et al. 2013, 
ApJ, 763, 41,

\bibitem[Counselman(1973)]{counselman} 
Counselman, C. C. 1973, ApJ, 180, 307 

\bibitem[Darwin(1879)]{darwin1} 
Darwin, G. H. 1879, The Observatory, 3, 79

\bibitem[Darwin(1880)]{darwin2} 
Darwin, G. H. 1880, Phil. Trans. R. Soc. A, 171, 713 

\bibitem[Fabrycky et al.(2014)]{fabrycky}
Fabrycky, D. C., Lissauer, J. J., Ragozzine, D., et al. 2014, ApJ, 790, 146 

\bibitem[Fang \& Margot(2012)]{fangmargot} 
Fang, J., \& Margot, J.-L. 2012, ApJ, 761, 92

\bibitem[Feng et al.(2020)]{feng}
Feng, F., Butler, R. P., Shectman, S. A., et al. 2020, 
ApJS, 246, 11, arXiv:2001.02577

\bibitem[Fulton et al.(2017)]{fulton} 
Fulton, B. J., Petigura, E. A., Howard, A. W. et al. 2017, 
AJ, 154, 109 

\bibitem[Galilei(1610)]{galileo} 
Galilei, G. 1610, Sidereus Nuncius; Translated by A. Van Helden 1989,
(Chicago: Univ. Chicago Press) 

\bibitem[Gillon et al.(2016)]{gillon} 
Gillon, M., Jehin, E., Lederer, S. M., et al. 2016, Nature, 533, 221 

%\bibitem[Gillon et al.(2016)]{gillon} 
%Gillon, M., Jehin, E., Lederer, S. M., Delrez, L., De Wit, J.,
%Burdanov, A., Van Grootel, V., Burgasser, A. J., Triaud, A.H.M.J.,
%Opitom, C., Demory, B.-O., Sahu, D. K., Bardalez Gagliuffi, D.,
%Magain, P., \& Queloz, D. 2016, Nature, 533, 221 

%\bibitem[Gomes et al.(2005)]{gomes} 
%Gomes, R., Levison, H. F., Tsiganis, K., \& Morbidelli, A. 
%2005, Nature, 435, 466

\bibitem[Hesse(1872)]{hesse} 
Hesse, L. O. 1872, Die Determinanten elementar behandelt (Leipzig)

\bibitem[Hidalgo et al.(2020)]{hidalgo} 
Hidalgo, D., Pall{\'e}, E., Alonso, R., et al. 2020, 
A\&A, in press, arXiv:2002.01755

\bibitem[Hut(1980)]{hut1980}
Hut, P. 1980, A\&A, 92, 167

\bibitem[Kokubo \& Ida(1995)]{kokida95} 
Kokubo, E., \& Ida, S. 1995, Icarus, 114, 247

\bibitem[Kokubo \& Ida(1998)]{kokida98} 
Kokubo, E., \& Ida, S. 1998, Icarus, 131, 171

\bibitem[Levrard et al.(2009)]{levrard} 
Levrard, B., Winisdoerffer, C., \& Chabrier, G. 2009, ApJ, 692, 9

\bibitem[Mayor et al.(2011)]{massdist} 
Mayor, M., Marmier, M., Lovis, C. et al. 2011, arXiv:1109.2497

\bibitem[Millholland et al.(2017)]{millholland} 
Millholland, S., Wang, S., \& Laughlin, G. 2017, ApJ Letters, 849, L33

\bibitem[Mills et al.(2019)]{mills2019} 
Mills, S. M., Howard, A. W., Petigura, E. A., et al. 2019, AJ, 157, 198 

\bibitem[Murray \& Dermott(1999)]{md}
Murray, C. D., \& Dermott, S. F. 1999, Solar System Dynamics 
(Cambridge: Cambridge Univ. Press) 

\bibitem[Nielsen et al.(2020)]{nielsen} 
Nielsen, L.D., Gandolfi, D., Armstrong, D. J., et al. 2020, 
MNRAS, in press, arXiv:2001.08834

\bibitem[Peale(1976)]{peale76}
Peale, S. H. 1976, ARA\&A, 14, 215 

\bibitem[Petit et al.(2018)]{petit} 
Petit, A., Laskar, J., \& Bou{\'e}, G. 2018, A\&A, 617, 93 

\bibitem[Pu \& Wu(2015)]{puwu} 
Pu, B., \& Wu, Y. 2015, ApJ, 807, 44

\bibitem[Rodriguez et al.(2020)]{rodriguez}
Rodriguez, J. E., Vanderburg, A., Zieba, S., et al. 2020, 
submitted to AAS Journals, arXiv:2001.00954

\bibitem[Rowe et al.(2014)]{rowe} 
Rowe, J. F., Bryson, S. T., Marcy, G. W., et al. 2014, ApJ, 784, 45

\bibitem[Steffen \& Hwang(2015)]{steffen} 
Steffen, J. H., \& Hwang, J. A. 2015, MNRAS, 448, 1956 

\bibitem[Strogatz(1994)]{strogatz} 
Strogatz, S. H. 1994, Nonlinear Dynamics and Chaos 
%-- with Applications to Physics, Biology, Chemistry, and Engineering
(Reading: Perseus Books) 

\bibitem[Tanaka \& Ward(2004)]{tanakaward} 
Tanaka, H., \& Ward, W. R. 2004, ApJ, 602, 388 

\bibitem[Tremaine(2015)]{tremaine15} 
Tremaine, S. 2015, ApJ, 807, 157

\bibitem[Tremaine \& Dong(2012)]{tredong}
Tremaine, S., \& Dong, S. 2012, AJ, 143, 94 

\bibitem[Tsiganis et al.(2005)]{tsiganis} 
Tsiganis, K., Gomes, R., Morbidelli, A., \& Levison, H. F. 
2005, Nature, 435, 459

\bibitem[Van Eylen \& Albrecht(2015)]{vaneylen}
Van Eylen, V., \& Albrecht, S. 2015, ApJ, 808, 126

\bibitem[Wang(2017)]{songhu}
Wang, S. 2017, RNAAS, 1, 26 

\bibitem[Weiss et al.(2018a)]{weissa} 
Weiss, L. M., Marcy, G. W., Petigura, E. A., et al. 2018a, AJ, 155, 48

\bibitem[Weiss et al.(2018b)]{weissb} 
Weiss, L. M., Marcy, G. W., Petigura, E. A., et al. 2018b, AJ, 156, 254

\bibitem[Weiss \& Petigura(2019)]{weisspet} 
Weiss, L. M., \& Petigura, E. A. 2019, arXiv:1908.05833 

\bibitem[Wolfgang et al.(2016)]{wolfgang} 
Wolfgang, A., Rogers, L. A., \& Ford, E. B. 2016, ApJ, 825, 19 

\bibitem[Zhu et al.(2018)]{zhu} 
Zhu, W., Petrovich, C., Wu, Y., Dong., S., \& Xie, J. 2018, 
ApJ, 860, 101

\end{thebibliography}
\end{document}